\documentclass[12pt,preprint]{aastex}

\def\lsim{\hbox{ \rlap{\raise 0.425ex\hbox{$<$}}\lower 0.65ex\hbox{$\sim$} }}
\def\gsim{\hbox{ \rlap{\raise 0.425ex\hbox{$>$}}\lower 0.65ex\hbox{$\sim$} }}

\def\opeqn{\begin{equation}}
\def\cleqn{\end{equation}}

\begin{document}

\title{ Reflection Symmetry of Cusps in Gravitational Lensing }
\author{Sun Hong Rhie (UND)}


\begin{abstract}
Criticality in graviational microlensing is an everyday issue because that is
what generates microlensing signals which may be of photon-challenged compact 
objects such as black holes or planetary systems ET calls home. 
The criticality of these quasi-analytic lenses is intrinsically quadratic, 
and the critical curve behaves as a mirror generating 
two mirror images along the image line ($\parallel \pm E_-$) at
the same distances from the critical curve in the opposite sides. 

At the (pre)cusps where the caustic curve ``reflects" and develops cusps, 
however, ``would-be" two pairs of quadratic images ``superpose" to produce 
three mirror images because of the degenerate criticality.   
The critical curve behaves as a parabolic mirror, and the image inside 
the parabola is indeed a superposed image having the sum of the magnifications 
of the other two that are outside the parabola. All three images lie on 
a parabolic image curve shaped by a property ($\nabla J$) of the (pre)cusp,
and {\it two distances} of the system determine the position
of the image curve and positions of the three images on the image curve.   
The triplet images satisfy $\sum J^{-1} =0$, and the function
$\sum J^{-1}$ is discontinuous at a caustic crossing
where a pair of quadratic images disappear into a critical point.
The reflection symmetry of the image curve is a manifestation of the symmetry
of the cusp which is also respected by a trio of parabolic curves that are 
tangnet at the (pre)cusp and define the image domains. 
The symmetry is guaranteed when $J_{+-}$ vanishes 
or can be ignored, and the cusps on the lens axis of the binary lenses are 
strongly symmetric having $J_{+-}=0$ 
because of the global reflection symmetry of the binary lenses. 
``$E_\pm$-algebra" is laid out for users' convenience.
\end{abstract}


\subsection*{ 0. \ \  {\it `` Forget About This Universe"} }

\begin{quotation}
\noindent
{\it Q: ~Quazi-What lenses? Too special! What do you think is the 
probability that one of the baby universes will evolve into one just like 
this?  Measure zero. \ A: ~Well, sir, one resolution is to preach a paradox. 
The other is to admit that you have not found the proper measure yet. 
-- ``Modo de Notre Dame"}
\end{quotation}

\clearpage

\section{ Stationary Points of the Critical Curve and Cusps}  

The simplest and well-known cusp is the point caustic of a single lens. The 
critical curve of the single lens is a circle better known as Einstein image ring.
A circle is locally parabolic in a small neighborhood of a point on the circle. 
If the circle is given by $u^2 + v^2 = a^2$, and $(u, v) = (a, 0)$ is a point
on the circle, then the equation of the circle becomes $2 a t \approx v^2$ 
where $ t = a - u$ and $v$ is sufficiently small. 
\opeqn   
 2 a t = v^2 + t^2  
\cleqn 
The criticality of a single lens is highly degenerate as is shown in equation
(\ref{eqPcaustic}). In fact, the eigenvector $E_+$ is radial everywhere, and 
$J_-$ and higher derivatives vanish. Aside from the single lens, quasi-analytic 
lenses do have non-vanishing second derivative $J_{--}\neq 0$ at the cusps, and
its relevance to the triple images near cusps can be seen in equation 
(\ref{eqOff}). One may refer to SEF for the cases where $J_{--}=0$ with
non-vanishing higher order derivatives. 
Here we are interested in quasi-analytic lenses that
include $n$-point lenses and their variations such as constant shear 
(astro-ph/0103463).  

We need two equations to examine the degenerate critical behavior of the lens 
equation. One is the differential of the lens equation, 
and the other is the differential of the Jacobian determinant 
of the lens equation. Since the binary lens equation is a typical quasi-analytic
lens equation, we use it for a concrete book-keeping of the differential behavior.
\opeqn 
 \omega = z - {\epsilon_1\over \bar z_1} - {\epsilon_2\over \bar z_2} 
   \qquad ; \qquad z_j \equiv z - x_j : \ j = 1, 2
\label{eqLeq}
\cleqn
where $\omega$ and $z$ are the source and its image position, and $x_1$ and
$x_2$ are the lens positions. Then, the Jacobian determinant is given by
\opeqn
 J = 1 - |\kappa|^2 \quad ; \qquad \kappa 
  \equiv \sum_j {\epsilon_j\over z_j^2} \ .  
\cleqn
Now, we differentiate the two equations. 
\begin{eqnarray} 
 d\omega &=& 
     dz + \bar\kappa d\bar z + {1\over 2} \bar\partial\bar\kappa~ d\bar z^2
        + {1\over 6} \bar\partial^2\bar\kappa~ d\bar z^3 + {\cal O}(dz^3)  
      \ \equiv \ \delta\omega_1 + \delta\omega_2 + \delta\omega_3 
\label{eqDLens}   \\
 dJ &=& (dz\partial + d\bar z \bar \partial) J +
   {1\over 2} (dz\partial + d\bar z \bar \partial)^2 J + {\cal O}(dz^2) 
    \  \equiv \   \delta J_1 + \delta J_2
\label{eqOne}
\end{eqnarray}
Criticality refers to the fact that the linear terms of equation (\ref{eqDLens}) 
can vanish, and two concepts need to be distinguished. One is the critical
curve where the criticality occurs in the image plane ($z$-plane), and the other
is the critical direction ($\pm E_-$) at each (critical) point on the critical
curve. The tangent of the critical curve at a critical point and critical 
direction at the same critical point generally are not the same. The angle
between the tangent and critical vector varies slowly along the critical curve,
and the caustic develops a cusp when the two coincide. The cusps are the 
interest here. First, we run through the algebraic machinery of the 
criticality of quasi-analytic lenses. 
\begin{enumerate}
\item
The Jacobian matrix of the lens equation is an array of the linear coefficients 
of equation (\ref{eqDLens}) and its complex conjugate, and $J$ is 
the determinant of the matrix.  
\opeqn
   {\cal J} = \pmatrix{1 \  \bar\kappa \cr
                       \kappa \  1 } \ ; \qquad J = det({\cal J})
\label{eqJacobian}
\cleqn
The analytic function $\kappa(z)$ completely determines the differential behavior,
and we denote its phase angle by $2\varphi$.
\opeqn
 \kappa = |\kappa| e^{i2\varphi}
\cleqn
\item
The eigenvalues of the matrix (\ref{eqJacobian}) are given by
\opeqn
 \lambda_\pm = 1 \pm |\kappa| \ . 
\cleqn
When one ($\lambda_-$) of them vanishes, the Jacobian determinant $J$ vanishes, 
and the lens equation becomes quadratic (or degenerate) in that direction 
($\pm E_-$). The critical condition ($J=0 \Leftrightarrow |\kappa|=1$) imposes
one real constraint on the otherwise two degrees of freedom of the 2-d $z$-plane,
and so the critical curve is a 1-d curve. We have emphasized 
that the critical direction is generally not the direction 
(or the tangent) of the critical curve. But, when they do coincide, 
the ``motion" along the critical curve becomes stationary, and
the curve in the $\omega$-space (caustic curve) develops a cusp. The critical
direction ($\pm E_-$) is the eigendirection of $\lambda_-$, and we can use the
orthogonal eigenvectors of the hermitian matrix (\ref{eqJacobian}) as an 
orthogonal basis vectors.
\opeqn
 e_\pm = {E_\pm \choose \bar E_\pm} \quad ; \qquad 
  E_+ = e^{-i\varphi} \ , \quad E_- = i E_+
\cleqn 
The linear terms of the local lens equation in (\ref{eqLeq}) reads as follows
where $dz = dz_+ E_+ + dz_- E_-$.
\opeqn
 \delta\omega_1 = dz + \bar\kappa d\bar z 
                = (1+|\kappa|) dz_+ E_+ + (1-|\kappa|) dz_- E_-
\cleqn  
At a critical point, $|\kappa|=1$, and the equation is stationary in the critical 
direction ($E_-$). If $z$ is a critical point, $\omega (z, \bar z)$ is called 
a caustic point, and the equation shows that the non-critical direction vector 
$E_+$ is tangent to the caustic curve: $\delta\omega_1 \parallel \pm E_+$.     
\item
In order to discuss higher order terms, we need to do some book-keeping for
orthogonal decompositions and differentiations in the bases of $E_\pm$. We  
note that $E_\pm (\varphi)$ are moving frame bases, or curvilinear
coordiante basis vectors, hence the corresponding tangent differential 
operators $\partial_\pm$ do not necessarily commute. 
\opeqn
 (\partial_+, \partial_-) 
 \equiv \partial_+ \partial_-  - \partial_- \partial_+ \neq 0
\cleqn  
The following is a list of notational definitions and useful relations. 
\begin{eqnarray*}
 && \partial = \partial/\partial z ; \qquad
      \bar\partial = \partial/\partial \bar z ;  \\
 && dz = dz_+ E_+ + dz_- E_- ;  \quad  
                  d\bar z = dz_+ \bar E_+ + dz_- \bar E_- ; \\  
 && dz_\pm   = {1\over 2} (\bar E_\pm dz  +  E_\pm d\bar z ) ;   \\ 
 && \delta\omega_k = \delta\omega_{k+} E_+ + \delta\omega_{k-} E_- ;  \\
 && \partial_\pm  = E_\pm \partial + \bar E_\pm \bar\partial  ;  \\  
 && \partial = {1\over 2} (\bar E_+ \partial_+  + \bar E_- \partial_-) ;  \quad 
         \bar\partial = {1\over 2} ( E_+ \partial_+  +  E_- \partial_-)  ;  \\
 && dz \partial + d\bar z \bar \partial 
     = dz_+ \partial_+ + dz_- \partial_-  ;  \\
 && dJ = \partial J dz + \bar\partial J \bar d\bar z = J_+ dz_+ + J_- dz_- ; \\
 && \bar\partial J = J_+ E_+ + J_- E_-  ;  \qquad
               \partial J = J_+ \bar E_+ + J_- \bar E_-   ; \\   
 && \partial E_\pm = {\partial J~ E_\pm \over 4|\kappa|^2} ; \qquad  
       \bar\partial E_\pm = - {\bar\partial J~ E_\pm \over 4|\kappa|^2} ;  \\
 && \partial_+ E_\pm = {\mp J_- E_\pm\over 4|\kappa|^2} ;  \qquad
       \partial_- E_\mp = {\mp J_+ E_\pm\over 4|\kappa|^2}  ;  \\ 
 && \partial_+ dz_\pm = {\mp J_- dz_\pm\over 4|\kappa|^2}  ;  \qquad
       \partial_- dz_\mp = {\mp J_+ dz_\pm\over 4|\kappa|^2}  ;  \\
 && (\partial_+,\ \partial_-)  
         \ \equiv \ \partial_+\ \partial_- - \partial_- \ \partial_+
           = ~2 i ~(\partial J \bar\partial - \bar\partial J \partial) ;  \\
 && (\partial_+, \partial_-)~ J = 0
         \quad \Rightarrow \qquad J_{+-} = J_{-+}  
\end{eqnarray*}
\item
For a single lens, $\kappa = z^{-2}$, where the lens is located at the origin.
If $z = r e^{i\theta}$, then $E_+ = e^{i\theta}$, and  
$\partial_+ = \partial_r$ and $\partial_- = r^{-1} \partial_\theta$. 
It should be clear why photometric lensing is considered a short-range 
phenomenon, why the Einstein ring size ($ = 1$ here) matters so much in lensing,
and why the caustic widths are so small and caustic crossings necessitate 
such frantic chases. 
\opeqn
 J = 1 - {1\over r^4} \ , \quad  
 J_+ = \partial_+ J = {4\over r^5} \ , \quad
 J_{++} = \partial_+^2 J = -{20\over r^6} 
\cleqn
And, also what the axial symmetry of the single lens does to the critical
behavior. \ 1. Every point on the critical curve is a precusp. 
\ 2. The gradient $\nabla J$ is radial everywhere. 
\ 3. The criticality is highly degenerate. 
\opeqn
 J_- = 0 \ , \qquad J_{+-} = 0 \ , \qquad J_{--}=0
\label{eqPcaustic}
\cleqn 
\item
\underline{\it Quadratic Mirror Images: \ \ }
We set $dz = dz_- E_-$ to see the behavior of the lens equation in the critical
direction. 
\begin{eqnarray} 
 && \delta\omega_2 = {1\over 2} \bar\partial \bar\kappa d\bar z^2
        = {1\over 2} \bar\partial J dz_-^2 \ ;   \\
 && J = dJ = \delta J_1 = J_+ dz_+ + J_- dz_- = J_- dz_-
\end{eqnarray}  
It is quadratic, and the critical direction is at an angle with the critical 
curve where $J_- \neq 0$. Speaking in terms of images, two images on the 
{\it image line} along the critical direction comes from a same source.
The {\it image line} is disected by the critical point, and the source 
is located in the direction of $\nabla J$ from the caustic point. 
Since $\nabla J$ (=$2 \bar\partial J$) does not vanish 
on the critical curve unless the critical curve has bifurcation points, the 
quadratic criticality is defined almost everywhere except where $J_-=0$.
At a (pre)cusp where $J_-=0$, the lens equation appears to be stuck on the 
critical curve ($dJ =0$), and that is an indication that there is a third image 
whose $J$-value vanishes (or its magnification diverges). Thus, the local lens 
equation (\ref{eqDLens}) becomes a cubic equation in the neighborhood of 
the (pre)cusp. 
The quadratic behavior of the lens equation and its failure near (pre)cusps
have been discussed for a quarter century and we also have detailed recently 
in astro-ph/0205067.  
\end{enumerate}

\section {Triple Mirror Images and Reflection Symmetry}

We calculate the second order dervatives of $J$ in order to examine 
the third order behavior of the lens equation (\ref{eqDLens}). 
For visual simplicity, we use the following denotations.       
\begin{eqnarray}
 && a \equiv J_+ \ , \qquad     b \equiv J_- \ ;  \\
 && a_+ \equiv J_{++} \ ,  \qquad a_- = b_+ \equiv J_{+-} \ ,  \qquad
        b_- \equiv J_{--} \  ;  \\ 
 &&  d\omega  = (x + i y) E_+ \ , \qquad   dz = (u + i v) E_+  \ . 
\end{eqnarray}
\begin{eqnarray}
 && \partial \bar \partial J = {1\over 4} \left( 
          (a_+ + b_-) + {1\over 4|\kappa|^2} (a^2 + b^2) \right) \ ;  \\ 
 && \bar\partial^2 J = {E_+^2\over 4} \left( 
      (a_+ - b_- + i 2 a_-) - {1\over 4|\kappa|^2} (a+ib)^2 \right)  \ ; \\
 && \delta J = \delta J_1 + \delta J_2 
    = (a u + b v) + {1\over 2} ( a_+ u^2 + a_- 2 u v + b_- v^2)
                    +{1\over 8} (ub-va)^2  \ .  
\end{eqnarray}
For the quasi-analytic lenses we are interested in here, $J = 1 - |\kappa|^2$, 
and  
\opeqn
 \partial \bar \partial J = -{|\partial J|^2\over |\kappa|^2} 
 \qquad \Rightarrow \qquad
 a_+ + b_- = -{5 (a^2 + b^2)\over 4 |\kappa|^2} \ < 0  \ . 
\cleqn
In the case of a single lens, $a_+ = -20$,  $b_-=0$, and $|\nabla J|=4$ on the 
critical curve ($r=1$).
Away from a cusp, $\delta J \approx \delta J_1$.  Near a cusp ($b=0$), $\delta J$ 
dominantly depends linearly on $u= dz_+$ and quadratically on $v = dz_-$.  
\opeqn
 \delta J = a u + {1\over 2} \left(b_- + {a^2\over 4} \right) v^2  
\cleqn
Therefore, the critical curve ($\delta J =0; \ J_\circ = 0$) is parabolic 
near a cusp.
\opeqn
  u = - {1\over 2a} \left(b_- + {a^2\over 4} \right) v^2
\cleqn 
Now we choose the cusp under consideration as the origin of the lens plane, 
and the eigendirection $E_+$ as the real axis so that  
$E_+ = 1 ~(\Leftrightarrow \varphi = 0)$. Then, $\bar\kappa_\circ = 1$,  
and the lens equation (\ref{eqDLens}) reads as follows.     
\begin{eqnarray}
 && \omega = z + \bar z + \alpha_2 \bar z^2 + \alpha_3 \bar z^3  \ ;  \\
 && \alpha_2 = -{a\over 4} \ ,   \quad
 \alpha_3 = {1\over 12} (b_- + {3a^2 \over 4 }) - {i~a_-\over 12}  \ . 
\label{eqCusp}
\end{eqnarray}
The coefficients are almost real.  In fact, we can ignore the imaginary
component of $\alpha_3$, ~$\Im(\alpha_3) = - a_-/12$, when we consider only 
the lowest order terms.  

When there is a reflection symmetry in the system, 
$a_- = b_+ = 0$ at the cusps on the symmtry axis. 
That is the case for the cusps on the lens axis of the binary lenses, as 
we can directly calculate from the binary equation (\ref{eqLeq}). Without
loss of generality, the real axis has been chosen to be the lens axis,
and $c.c$ stands for complex conjugate as usual. Then,
\opeqn
 b_+ = \partial_+ \left(- E_- \bar\kappa \partial\kappa + c.c \right) 
     = - E_- \partial_+ (\bar\kappa \partial\kappa) + c.c
     = - i \bar\kappa \partial^2\kappa + c.c = 0 \ . 
\cleqn
The last equality holds because 
$\bar\kappa \partial^2\kappa = \partial^2\kappa$ is real on the real axis.   
\opeqn
 \partial^2\kappa = \sum_j {\epsilon_j\over z_j^4} \ ; \qquad
 \bar\kappa = 1
\cleqn 
Thus, the cusps on the lens axis of the binary lenses have strong
reflection symmetries.

\subsection { A Source on the Symmetry Axis: \ $\omega = \bar\omega$ }

The lens equation is factorized for a source on the symmetry axis,
$\omega = \bar\omega$.  
\opeqn  
 0 = \omega - \bar\omega 
   = (z-\bar z) \left(\alpha_2(\bar z + z ) 
              + \alpha_3 (\bar z^2 + \bar z z + z^2) \right)
\cleqn
Therefore, there is one image on the symmetry axis ($z = \bar z$), and there can 
be two images off the symmetry axis depending on the position of the source.   
\opeqn
 0 = \alpha_2 ~2 u - \alpha_3 ~v^2  
\cleqn 
\begin{enumerate}
\item
 {\it One image on the symmetry axis: \ \ } 
\opeqn
 u_1 = {\omega \over 2 } = {x\over 2} \ , \quad v_1 = 0 \ ;  \qquad
 J_1 = au = {a \omega \over 2 }  \ . 
\cleqn
\item
 {\it Two images exist off the symmetry axis when $(a b_-)\omega < 0$: \ \ } 
\opeqn
 u_{2,3} = {\omega\over 2} \left(1 + {3a^2\over 4 b_-} \right) \ , \quad
 v_{2,3} = \pm \left( - {3 a \omega \over b_- } \right)^{1/2}  \ ; \qquad
 J_{2,3} = - a \omega  
\label{eqOff}
\cleqn 
The two images are twin images with the
same magnification and parity because of the reflection symmetry retained by
the source on the symmetry axis. The symmetry axis is mapped to a parabolic
curve (see figure \ref{fig-cusp1}).
\opeqn
  u_{2,3} = -{a\over 6} v_{2,3}^2 \left( {3\over 4} + {b_-\over a^2} \right)
\cleqn
\item
The three images satisfy $\sum J^{-1} = 0$. The image on the symmetry axis
has the same magnification as the sum of the magnifications of the images
off the symmetry axis: \ $|J_1|^{-1} = |J_2|^{-1} +  |J_3|^{-1}$. 
The off axis images have the same parity ($J_2 J_3 > 0$) which is the opposite
to the parity of the image on the symmetry axis 
($J_1 J_2 J_3 <0$ in figure \ref{fig-cusp1}).   
\item
The three images are on the parabolic curve (image curve) 
given by the following equation.
\opeqn
 u = u_1 - {a \over 8} v^2
\label{eqIcurve}
\cleqn 
The shape of the image curve is determined by $\nabla J = a E_+$ at the cusp, 
and the (vertex) position of the curve is determined by the source position 
$\omega$: ~$u_1 = \omega/2$.  The image curve intersects with the critical 
curve at $(u, v) = (u_c, v_c)$ when $(a b_-)\omega < 0$, or equivalently
when there are two images off the symmetry axis.   
\opeqn
 v_c = \pm \left(- {a w \over b_-} \right)^{1/2}  \quad ; \qquad
 u_c = {\omega\over 2} \left( 1 + {a^2\over 4 b_-} \right)  
\label{eqVcUc}
\cleqn
The corresponding caustic points $(x, y_c)$ can be found 
from equation (\ref{eqArbit}).  
\opeqn
 y_c = \pm \left(- {a w \over b_-} \right)^{1/2} {a\omega\over 6} \ .  
\label{eqYc}
\cleqn
One trivial nontheless useful observation is that the source position
satisfies $0= |y| \le |y_c|$.  
\end{enumerate}

\subsection{ An Arbitrary Source Near a Cusp }
 
\begin{figure}
\plotone{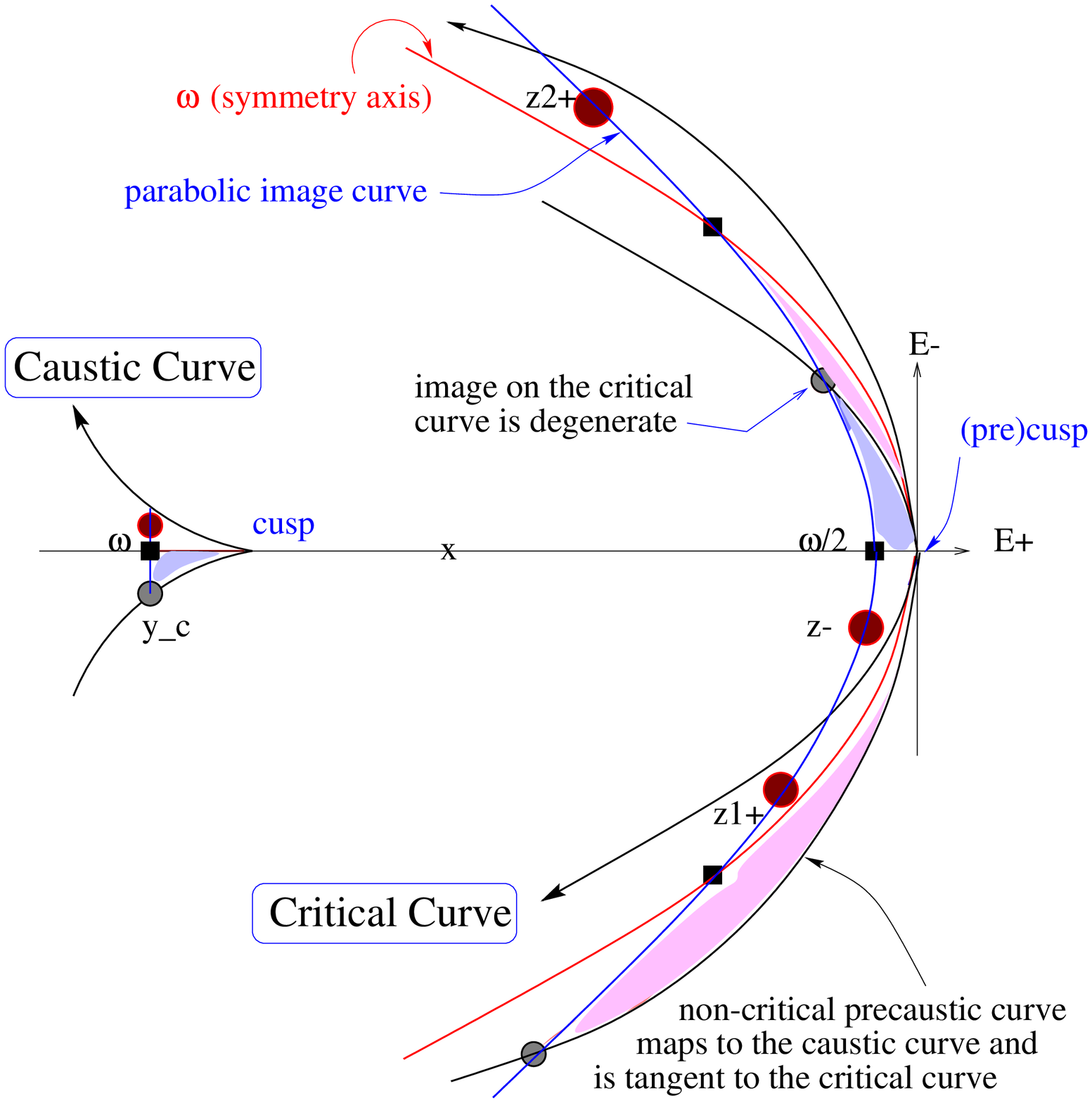}
\figcaption{\label{fig-cusp1} 
Three curves are tangent at the precusp: critical curve, a non-critical 
precaustic curve, and the image of the symmetry axis. Precaustic curves
in the image plane define image domains, and the critical curve is one of them.
The critical curve is tangent to a non-critical precaustic curve at a cusp
and the neighborhood of the precusp is divided into four image domains. 
The degeneracy is two fold and defines the reflection symmetry axis. 
The parabolic image curve at $\omega/2$ is the image of the 
source line at $\omega$ and the intersecion points with the non-critical 
precaustic curve define the range of the image curve. Three source positions 
and their images are shown to guide the eyes. The $1:3$ correspondence is 
shown for an area inside the caustic curve. Images $z_-$ and $z_{1+}$ resemble
a pair of quadratic images, but they are part of the triplet with $z_{2+}$, 
and the magnifications of $z_-$ and $z_{1+}$ are not the same.  
$A(z_-) = A(z_{1+}) + A(z_{2+})$.
}
\end{figure}

The images of an arbitrary source near a cusp can be discussed just as easily.  
In the lowest order approximation, where $u \sim v^2$, the lens equation 
(\ref{eqCusp}) near a cusp reads as follows. 
\opeqn  
 x = 2 u - \alpha_2 v^2 \quad ; \qquad
 y = - v (\alpha_2~ 2u - \alpha_3 v^2)
\label{eqArbit}
\cleqn
\begin{enumerate}
\item {\it Image Curve: \ }
The first equation is nothing but the parabolic image curve in equation 
(\ref{eqIcurve}) with the vertex position $\beta = x/2$ determined by the  
$E_+$-component $x$ of the source position. 
\opeqn
 u = {x\over 2} - {a \over 8} v^2
\cleqn
The images are on the parabolic image curve parameterized
by $x$ and shaped by the properties of the cusp -- reflection symmetry and
the value of $\nabla J$. Since the image curve is independent of $y$, an 
arbitrary source on the line $(x, y)$ where $x = 2\beta$ produces images 
on the same image curve determined by $x$.
The image curve intersects with the critical curve when $(a b_-) x < 0$.
\opeqn
 v_c = \pm \left(- {a x \over b_-} \right)^{1/2}  ; \quad
 u_c = {x\over 2} \left( 1 + {a^2\over 4 b_-} \right)  ; \quad
 y_c = v_c {a x\over 6} 
     = \pm \left(- {a x \over b_-} \right)^{1/2} {a x\over 6}
\cleqn
These are virtually identical equations with (\ref{eqVcUc}) and (\ref{eqYc}).
The source on the symmetry axis belongs to a family of sources that produce 
images on the image curve determined by the $E_+$-position of the source $x$.   
\item 
{\it The Number of Images: \ }
We eliminate $u$ from the two equations in (\ref{eqArbit}) and obtain
a standard third order real polynomial equation satisfied by $v$. 
\opeqn
 0 = v^3 + {3 a x\over b_-} v - {12 y\over b_-}  \ \equiv \ 
               v^3 + p v - q 
\label{eqCubic}
\cleqn 
So, there can be one, two, or three images on the image curve depending 
on the value of $y$.   
The number of real solutions of the cubic equation (\ref{eqCubic})
is related to the sign of the discriminant $D$: \ $\#  = 2 - sgn(D)$.
\opeqn
 D = \left( {ax\over b_-} \right)^3 + \left( {6y\over b_-} \right)^2
   = {36\over b_-^2} \left(y^2 - y_c^2 \right) \ ,
\cleqn
where
\opeqn
 y_c^2 \equiv - {a^3 x^3 \over 36 b_-} \ ; \qquad y_c^2~(a x b_-) < 0  \ .
\cleqn
If $(a x b_-) < 0$, then $y_c^2 > 0$ and the number of images is
three, two, or one depending on $|y| < |y_c|$, $|y| = |y_c|$,
or $|y| > |y_c|$.  If $(a x b_-) > 0$, then $D > 0$, and there is one image.
We recall that $(a b_-) x < 0$ is the condition that the image curve 
intersect with the critical curve. Since a source on the symmetry axis
($y = 0$) have three images, we can expect from continuity that sources 
with $|y| < |y_c|$ are inside the caustic and produce three images.  
It is consistent with the count of the solutions determined by $D$. 
\item  {\it The Third Image on the Precaustic Curve: \ }
The equation has the standard form of Vieta where the coefficient of the second 
order term vanishes. For our purpose, it means that $\sum_j v_j = 0$ where
$v_j: \ j = 1,2,3$ are the $E_-$ positions of the images two of which
can be complex.  For example, the images of a source at a caustic point
$(x, y_c)$ can be found trivially: A degenerate image at a critical point 
$(u_c, v_c)$ and the third image at $(u_3, v_3)$ where $v_3 = - 2 v_c$
($0 = \sum_j v_j = v_c + v_c + v_3$). 
\opeqn
 u_3 = {x\over 2} \left( 1 + {a^2\over b_-} \right) \ ; \qquad
 v_3 = -2 v_c \ . 
\cleqn   
The degenerate image is on the critical curve and so has $J = 0$. 
For the third image, 
\opeqn
 J  = {1\over 2} \left(a x + b_ - v_3^2 \right) = -{3 ax \over 2} \ . 
\cleqn 
\item 
{\it The Precaustic Curve Tangent to the Critical Curve at the PreCusp: \ } 
We recall that the critical curve is not the only
curve that is mapped onto the caustic curve under the lens equation
(see astroph.0103463.fig.10).
The curves in the image plane that are mapped onto the caustic curve
are called {\it precaustic curves} and the {\it precaustic curves}
contitute the boundaries of the image domains. The critical curve is
one of the precaustic curves, and that smooth one. The non-critical precaustic
curves are cuspy at the cusps of the caustic curve, except at the precusps 
where they are tangent to the critical curve.  
At a precusp, the critical curve and one non-critical precaustic curve
are tangent. In other words, the third image of the source $(x, y_c)$  on 
the caustic curve is on the non-critical precaustic
curve that is tangent to the critical curve at the precusp under
consideration (at the origin here). By eliminating the paramter $x$
from $u_3$ and $v_3$, we find that the precaustic curve is a parabolic 
curve with the vertex at the origin.  
\opeqn
 u_3 = -{a\over 8} v_3^2 \left(1 + {b_-\over a^2} \right)
\label{eqPrecaustic}
\cleqn 
If $|y| > |y_c|$, then $|v| > 2|v_c|$, and the third image is outside the 
parabola of the precaustic curve. See figure \ref{fig-cusp1}.  
The precaustic curve in equation (\ref{eqPrecaustic}) 
defines the boundary of the image domains that are mapped to inside the 
caustic and outside -- or the image domains of triplets and of singlets.
\item 
{\it Sum Rule \ $\sum J^{-1} = 0$ for Triple Images:  \ }  
We intuitively expect that the triple critical images near a precusp 
should satisfy $\sum J^{-1} = 0$ because they should be a ``superposition"
of the ``would-be" two pairs of quadratic images. A pair of quadratic images 
connected by the imageline bisected by the critical curve have the same
magnification and opposite parities, {\it i.e.,} $\sum_1^2 J^{-1}=0$.
In fact, the triplet images of a source on the symmetry axis satisfy
$\sum J^{-1}=0$. On the other hand, the images of a source on the
caustic curve fail to satifsy the relation $\sum J^{-1}=0$. Thus, we  
expect that the function $\sum J^{-1}$ is not smooth but discontinuous
at a caustic crossing. From equation (\ref{eqCubic}), we obtain
\begin{eqnarray}
 && J = {1\over 2} \left(ax + b_- v^2 \right) 
     = -ax \left(1 - {t\over v} \right) \ ; \qquad 
       t \equiv {6y\over ax}   \\
 \Rightarrow \ && 
 \sum J^{-1} = {-2P t + 3 Q \over \Pi_j (t - v_j)}  \ .  
\end{eqnarray}
The numerator vanishes. Therefore, the sum rule holds, and the total 
magnification of the positive images ($A_+$) is the same as the total 
magnification of the negative images ($A_-$).  
\opeqn 
 \sum J^{-1} = 0  \quad  \Rightarrow \quad  A_+ = A_- \ ;
  \qquad  \Pi_j (t- v_j) \neq 0
\cleqn
Since the ``middle image" has the opposite parity to the others, the 
magnification of the ``middle image" is the same as the total magnification
of the other two images. In figure \ref{fig-cusp1}, the ``middle image"
is the negative image, and that is always the case for the cusps on the 
symmetry axis of the binary lenses. 
\begin{enumerate}
\item
{\it Breakdown of the Sum Rule: \ }
We note that $t = v_c$ when $ y = y_c$. Thus, the denominator
$\Pi_j (t-v_j)$ vanishes for a source on the caustic, and $\sum J^{-1}=0/0$ 
is indeterminate. In other words, the sum rule for the triplet breaks down
at a caustic crossing. 
\end{enumerate}
\item
{\it Decoupling of the Triplet to Doublet $\oplus$ Singlet: \ } 
If we consider that the two images 
disappearing into the critical curve are a pair of quadratic images, it should
be clear that the sum rule $\sum_{doublet} J^{-1}=0$ holds between the two 
quadratic images, and it is impossible to satisfy the sum rule for the 
triplet. For practical purpose, the transition from a {\it triplet} to 
{\it doublet $\oplus$ singlet} occurs where the magnification of the
``middle image" becomes much larger than that of the third image
$|J^{-1}| (u_3, v_3) = |2/3ax|$. We can consider the inverse transition to
be where the behavior of a pair of images  near the critical curve
fails to be of a quadratic mirror image pair. Since the
magnification of the third image is inversely proportional to $|x/2|$, 
the (vertex) position of the image curve, the validity range of the 
quadratic images converges to zero as the source nears the cusp.
In figure \ref{fig-cusp1}, as the two 
images $z_-$ and $z_{1+}$ move toward the critical curve, $z_{2+}$     
moves toward the non-critical caustic curve, and the magnifications 
of the $z_-$ and $z_{1+}$ become almost equal. 
\item
{\it Magnifications of the Three Images: \ } 
The sum rule $\sum J^{-1}$ is an analytic topological quantity, and we only had
to manipulate the cubic equation without solving it.  
The magnification of the individual images requires the solutions, and the 
solutions of the cubic equation (\ref{eqCubic})     
can be found in mathematics encyclopedia such as by E. Weisstein (1999).
The trick is simple. Start with the cubic equation $0 = v^3 + p v - q$.
Substitute $v$ by $ \sigma - p/ 3\sigma$ (Vieta) and solve the quadratic 
equation for $\sigma^3$.  For three images, $D < 0$.
\opeqn
 v = \sigma - {p\over 3\sigma} = \sigma + \bar\sigma = 2 |\sigma| \cos\theta  
 \ ; \qquad  |\sigma| = \left(-{p\over 3} \right)^{1/3} 
                      = \left(-{ax\over b_-} \right)^{1/3}  , 
\cleqn
and $\theta$ can be (numerically easily) found from the following.   
\opeqn
  \theta = {1\over 3}tan^{-1} \left( {2|D|^{1/2}\over q} \right)
\ \ \Leftarrow \ \ 
 |\sigma|^3 \cos 3\theta = {q\over 2} \ ; \ \ 
 |\sigma|^3 \sin 3\theta = |D|^{1/2}   
\cleqn
Then, the three solutions are $\theta$, and $\theta \pm 2\pi/3$.
The Jacobian determinant follows easily.
\opeqn
  J = {1\over 2} \left( ax + b_- v^2 \right) 
    = -ax + {6y\over 2|\sigma|\cos\theta} \ (v \neq 0)
\label{eqTripletJ}
\cleqn  
If $y=0$, then $\theta = \pi/6$, and we recover the solutions of a source
on the symmetry axis: \  $v = 0, \pm\sqrt{3}|\sigma|$; \ $J = ax/2, -ax$ 
where the latter two exist when $(ab_-) x < 0$.  
For $y\neq 0$, the algebraic expressions may not be so illuminating than figures. 
Figures \ref{fig-positive} and \ref{fig-negative} show the general behavior
of the positive and negative cusps of a binary lens caustic. The cusp
at $\omega = 0.757$ shows a typical orderly distribution of the iso-$J$ curves
and how the multiplicity of images occurs. The cusp $\omega = 0.757$ is a 
positive cusp and each positive iso-$J$ curve self-intersects inside the cusp.
Thus, there is only one positive image outside and two positive images inside.
Figure \ref{fig-negative} shows that there is one negative image and none
outside.  Being inside the binary lens caustic, 
the source should produce five images (two positive and three negative). 
The other two images are negative fainter images near the two lens positions
and are not represented in the figures because the $J$-value is cut at 
$J \sim -1.0$. The images at the lens positions have $J = -\infty$. 
The three bright images are the ones that are obtained as three solutions of 
the cubic equation when the source is near the cusp. The iso-$J$ contours can 
be found from equation (\ref{eqTripletJ}). Also, see figure 4 in astro-ph/0206162 
(Gaudi and Petters).           
\item
{\it Singlet Images Outside the Non-critical Precaustic Parabola: \ }       
There is one image when $D > 0$, and the image and its magnification can
be found using the same procedure.  
\begin{eqnarray}
 && v = \sigma - {p\over 3\sigma} = \sigma_+ + \sigma_-  \ ;  \qquad
 \sigma_\pm = \left( {q\over 2} \pm \sqrt{D} \right)^{1/3} \\
 && J = {1\over 2} \left( ax + b_- (\sigma_+ + \sigma_-)^2 \right) 
\end{eqnarray} 
If $x=0$, then $p = 0$ and $v= q^{1/3}$. Therefore, $J \propto y^{2/3}$.
\opeqn
 J =  \left(18 b_- y^2\right)^{1/3}
\cleqn
 \begin{enumerate}
  \item 
  The parity of the singlet image outside the non-critical precaustic parabola
is determined by the sign of $b_-$. A cusp is referred to as {\it a positive cusp}
if $b_- >0$ and {\it a negative cusp} if $b_- <0$. The ``superposed image" of 
the triplet near a positive (negative) cusp is a negative (positive) image. 
 \item
 Figures \ref{fig-positive}, \ref{fig-negative}, and \ref{fig-last} show
the six-cusped connected caustic with lens parameters 
$\epsilon_2=0.1$ and separation $\ell =1.3$. The cusps on the lens axis are
positive cusps and very strong. The cusps off the lens axis are negative cusps.
The negative cusps nearer to the smaller are weaker. The cusp near the larger 
mass is very strong and mimicks the behavior of a single lens producing
iso-$J$ curves (or ``balloons") that are almost circular (\ref{fig-positive}). 
  \item
 Given a value of $J$, $y \propto |b_-|^{1/2}$. Thus, the larger is $|b_-|$,  
 the tighter are the contours of iso-$J$. The iso-$J$ curves tend to be oblong
 in the direction of symmetry axis because $ y \sim J^{3/2}$ and $x \sim J$.
 \end{enumerate}
\item
{\it The Position of the Cusp: \ } Let's calculate the position of the cusp
(and precusp) on the lens axis near the smaller mass in 
figure \ref{fig-positive}. The position of the precusp in the image plane can
be found by sovling the following quartic equation where $\epsilon_2=0.1$ 
and $|x_2-x_1|=\ell$.     
\opeqn
 1 = {\epsilon_1\over (z-x_1)^2} + {\epsilon_2\over (z-x_2)^2} \ ; \qquad 
 z = \bar z
\cleqn
When caustic curve is connected, there are two real solutions, and the one
near the smaller mass is at $z = 1.553$ and the corresponding caustic position
$\omega = 0.759$ (in the center of mass coordinate system).  
\item
Figure \ref{fig-last} shows an example of ``perfectly good" line caustics
that are by definition clear of cusp ``balloons".
Dotted curves are $J = 0.88$ and $0.9$
and tracing their variations requires good quality photometry available
only from space microlensing. For ground-based alert networks, such line
caustic crossing will be without warning unless it is the exit crossing.
It is necessary to pin down the lens parameters in order to utilize the exit
line caustic crossing for limb darkening measurements, and that requires
sufficient information of the entry caustic crossing, which in turn requires
a reasonable time resolution survey ($\gsim 2 /day$) and immediate follow-up
capabilities. In addition, better than $1\%$ photometry is required to be able
to discern the linear limb darkening parameter by $10\%$ (rb99).
\end{enumerate}

\def\aj{\ref@jnl{AJ}}
\def\apj{\ref@jnl{ApJ}}
\def\apjl{\ref@jnl{ApJ}}
\def\apjs{\ref@jnl{ApJS}}
\def\aap{\ref@jnl{A\&A}}
\def\aapr{\ref@jnl{A\&A~Rev.}}
\def\aaps{\ref@jnl{A\&AS}}
\def\mnras{\ref@jnl{MNRAS}}
\def\prl{\ref@jnl{Phys.~Rev.~Lett.}}
\def\pasp{\ref@jnl{PASP}}
\def\nat{\ref@jnl{Nature}}
\def\iauc{\ref@jnl{IAU~Circ.}}
\def\aplett{\ref@jnl{Astrophys.~Lett.}}
\def\annrev{\ref@jnl{Ann.~Rev.~Astron.~and Astroph.}}

\clearpage


%
%

\begin{figure}
\plotone{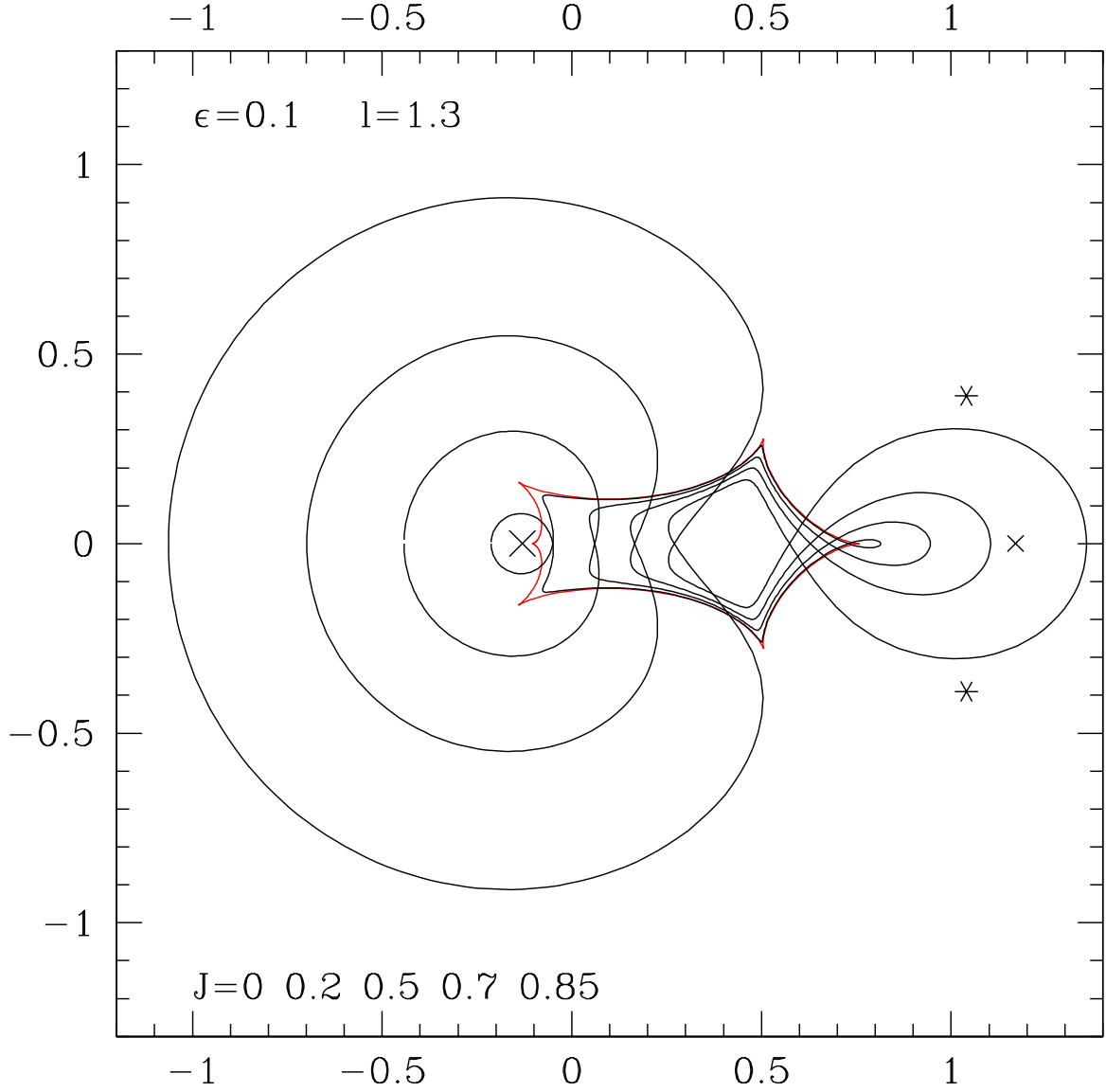}
\figcaption{\label{fig-positive}
Contours of $J =$ constant for $J \ge 0$. 
The two cusps on the lens axis are positive cusps.
The cusp at $\omega = 1.553$ shows the typical distribution of the $J$-curve
distribution near it. The cusp near the larger mass shows the same behavior
but for much smaller $J$-values not shown here. Near the cusps, the iso-$J$ 
curves can be obtained by solving the cubic equation. 
}
\end{figure}

\begin{figure}
\plotone{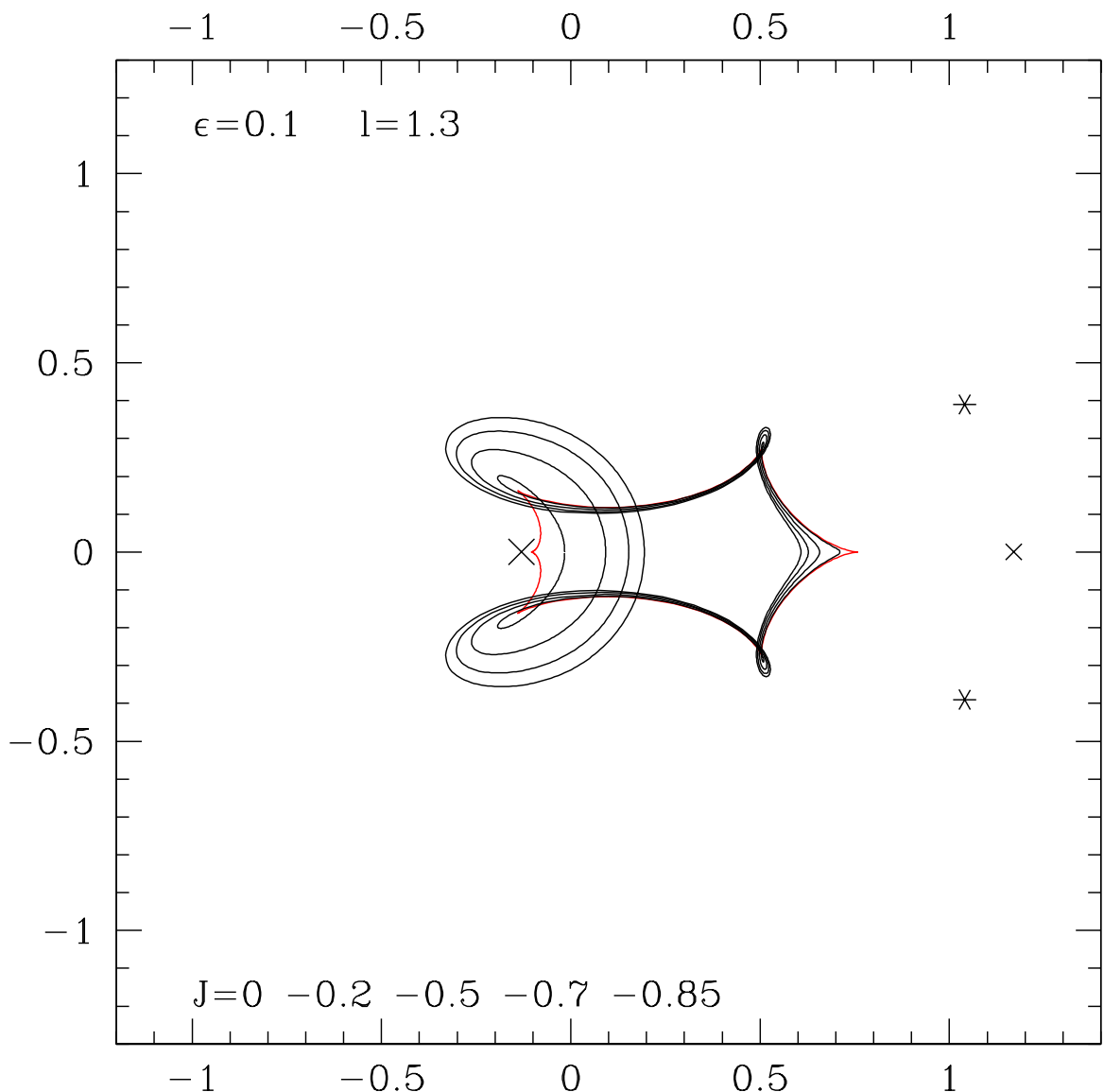}
\figcaption{\label{fig-negative}
Contours of $J =$ constant for $J \le 0$.
The four cusps off the lens axis are negative cusps.
The distribution of the negative iso-$J$ contours is compact in comparison
to the positive contours. Negative values $J = (-\infty, 0)$ are confined to 
the finite area inside the critical curve while the positive values are
distributed over the entire image plane outside the critical curve, and the
steep gradient of the distribution of negative iso-$J$ contours should be
expected. The curves neatly lining the caustic curve from inside represent
the ``widths" of the line caustics.     
}
\end{figure}

\begin{figure}
\plotone{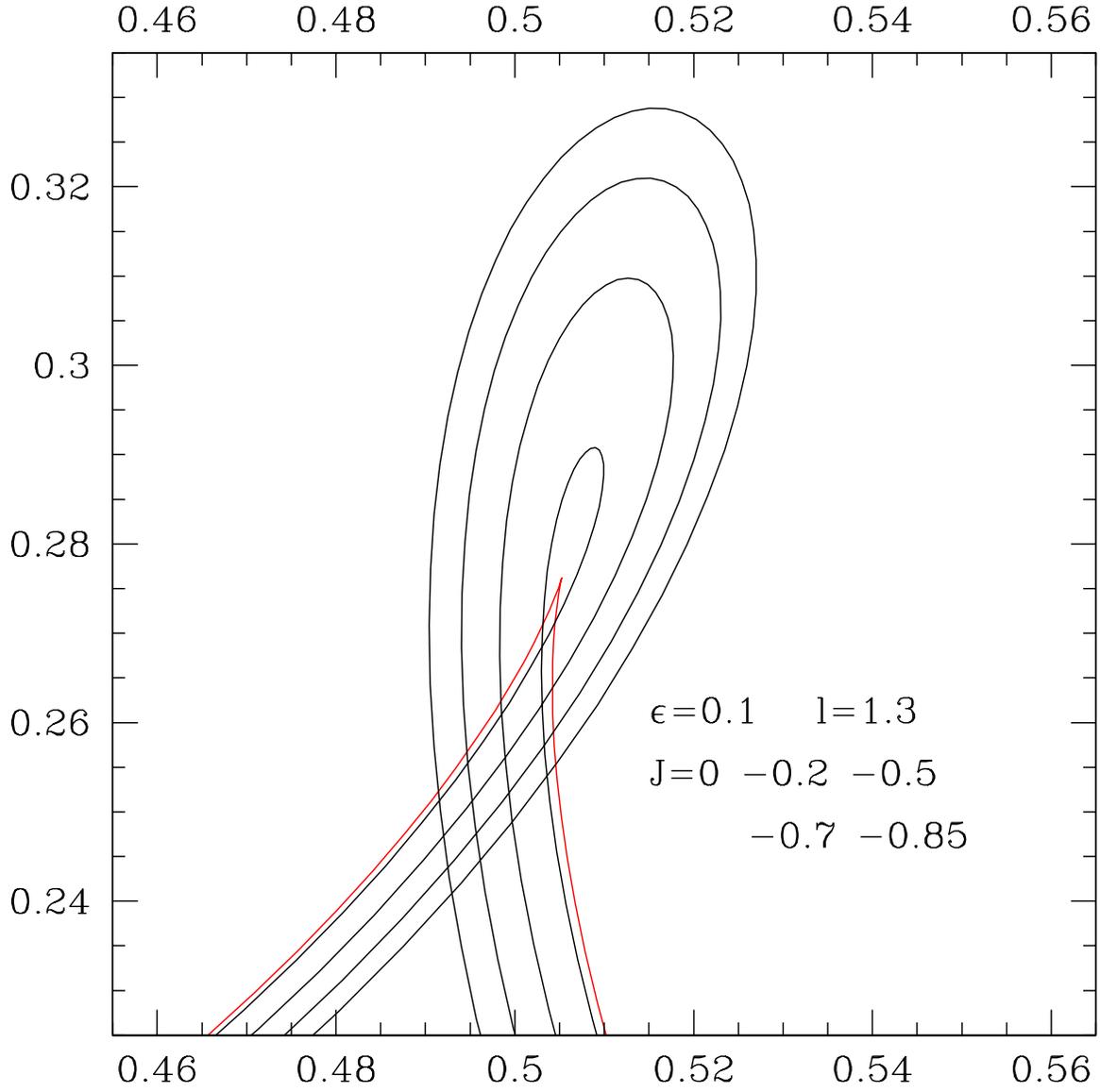}
\figcaption{\label{fig-offcusp}
Blow-up of the negative cusps nearer to the smaller mass.
}
\end{figure} 

\begin{figure}
\plotone{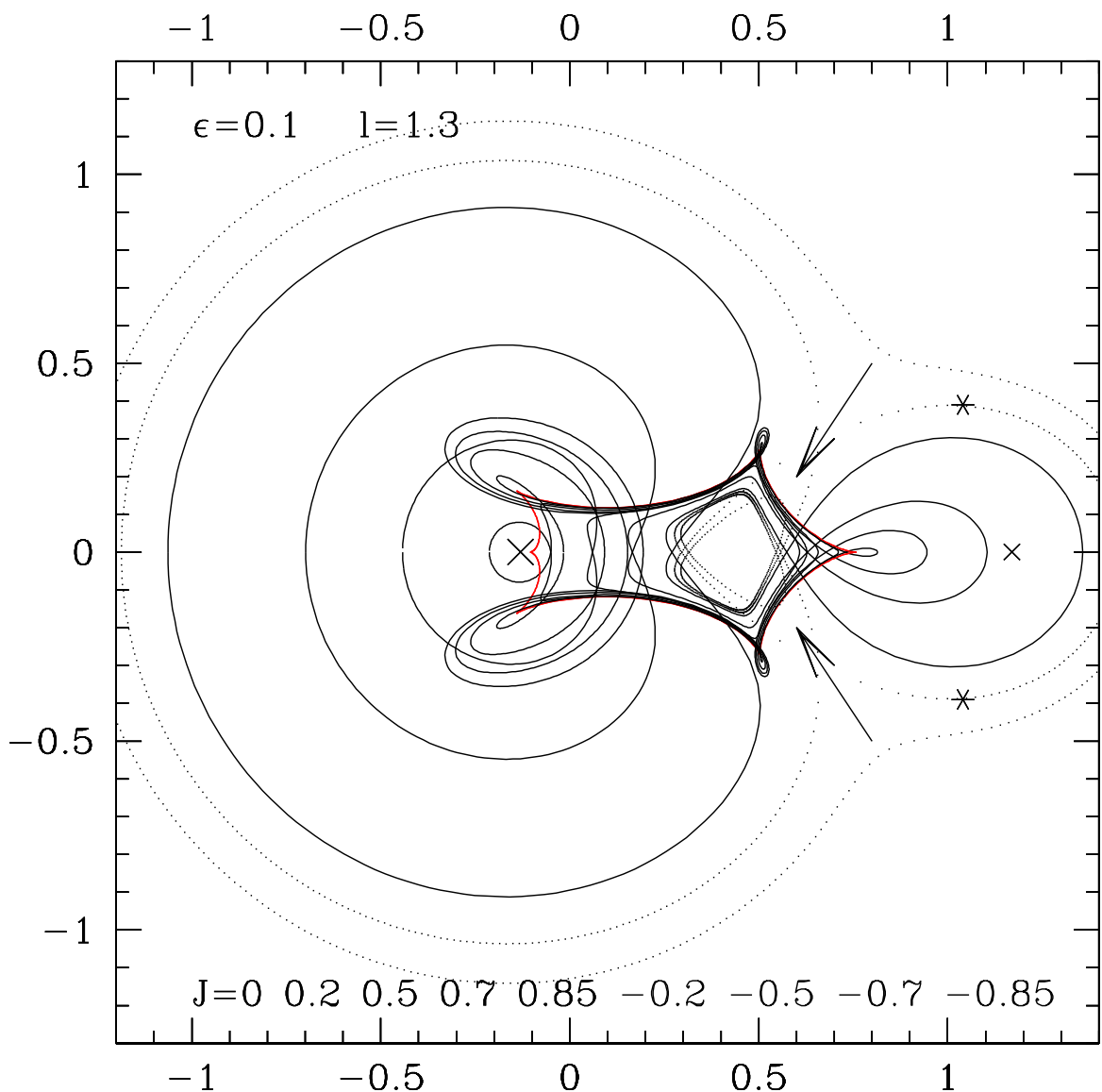}
\figcaption{\label{fig-last}
Arrows point at examples of ``prefectly good" line caustics that are by 
definition clear of cusp ``balloons". Dotted curves are $J = 0.88$ and $0.9$.
The area clear of iso-$J$ curves centered around $\sim (0.45, 0)$ is the 
location of the lowest magnifications inside the caustic and is responsible
for glacier-carved $U$-shape valley in the light curves. That is also typical
of diamond shape four-cusped caustics.  Five images of a source 
inside the caustic can have the absolute minimum magnification 3 when the separation 
$\ell = \sqrt{2}$. In that case, two of the five images are at the so-called
finite limit points marked by $\ast$'s.} 
\end{figure}

\end{document}